\documentclass[aps,prl,reprint,superscriptaddress]{revtex4-1}
\usepackage{amsmath}
\usepackage{graphicx}
\usepackage{amsfonts}
\usepackage{color}
\usepackage{upgreek}

\usepackage[table]{xcolor}
\usepackage{textcomp}
\usepackage{amsmath}
\usepackage{amssymb}
\usepackage{graphicx} 
\usepackage{pgf}
\usepackage{epstopdf}
\usepackage{braket}
\usepackage{mathtools}

\usepackage{soul}

\def\bra#1{\langle #1 |}
\def\ket#1{| #1 \rangle}

\definecolor{islamicgreen}{rgb}{0.0, 0.56, 0.0}

\usepackage[colorlinks,linkcolor=blue,urlcolor=blue,citecolor=blue]{hyperref}
\begin{document}

\title{Multipartite entanglement structure in the Eigenstate Thermalization Hypothesis}

\author{Marlon Brenes}
\email{brenesnm@tcd.ie}
\affiliation{Department of Physics, Trinity College Dublin, Dublin 2, Ireland}

\author{Silvia Pappalardi}
\email{spappala@sissa.it}
\affiliation{SISSA, Via Bonomea 265, I-34135 Trieste, Italy}
\affiliation{ICTP, Strada Costiera 11, I-34151 Trieste, Italy}

\author{John Goold}
\affiliation{Department of Physics, Trinity College Dublin, Dublin 2, Ireland}

\author{Alessandro Silva}
\affiliation{SISSA, Via Bonomea 265, I-34135 Trieste, Italy}
  
\date{\today}

\begin{abstract}
We study the quantum Fisher information (QFI)  and, thus, the multipartite entanglement structure of thermal pure states in the context of the Eigenstate Thermalization Hypothesis (ETH). In both the canonical ensemble and the ETH, the quantum Fisher information may be explicitly calculated from the response functions. In the case of ETH, we find that the expression of the QFI bounds the corresponding canonical expression from above. This implies that although average values and fluctuations of local observables are indistinguishable from their canonical counterpart, the entanglement structure of the state is starkly different; with the difference amplified, e.g., in the proximity of a thermal phase transition. We also provide a state-of-the-art numerical example of a situation where the quantum Fisher information in a quantum many-body system is extensive while the corresponding quantity in the canonical ensemble vanishes. Our findings have direct relevance for the entanglement structure in the asymptotic states of quenched many-body dynamics. 
\end{abstract}

\maketitle
\emph{Introduction}.--- Thermalization is a phenomenon in many-body physics that occurs with a high degree of universality~\cite{Gallavotti:2013}. The question of how and why thermalization emerges from unitary quantum time evolution was posed even in the inception of quantum theory by some of its founding fathers~\cite{Schrodinger:1927,Vonneumann:1929,Goldstein:2010}. Nature shows us that the evolution of a pure, thermally isolated system typically results in an asymptotic state that is indistinguishable from a finite temperature Gibbs ensemble by either local or linear response measurements
. 
One predictive framework for understanding thermalization from quantum dynamics is the Eigenstate Thermalization Hypothesis (ETH). 
Inspired by early works by Berry~\cite{Berry:1977,Berry:1977b}, later formulated by Deutsch~\cite{Deutsch:1991}, ETH was fully established by Srednicki as a condition on matrix elements of generic operators $\hat O$ in the energy eigenbasis~\cite{Srednicki:1994,Srednicki:1996,Srednicki:1999}. Subsequently, ETH has motivated a considerable body of numerical work over the past decade~\cite{Rigol:2008,Polkovnikov:2011,Alessio:2016}.
Far from being an academic issue, thermalization in closed quantum systems is now regularly scrutinized in laboratories worldwide where 
advances in the field of ultra-cold atom physics have allowed for probing quantum dynamics on unprecedented timescales in condensed matter physics
~\cite{Kinoshita:2006,Lewenstein:2007,Polkovnikov:2011,Bloch:2012}.

Whenever the ETH is satisfied, it is difficult to contrast the coherence of a pure state with that of a statistical mixture by means of standard measurments. Therefore, a question that naturally comes to mind is: will pure state dynamics possess detectable features beyond thermal noise? This question, posed recently by Kitaev~\cite{kitaevTalk} in the context of black-hole physics, lead him to suggest the study of
a peculiar type of out-of-time-order correlations (OTOC), originally introduced by Larkin and Ovchinikov~\cite{larkin1969quasiclassical}. This object, as a result of a nested time structure, detects quantum chaos and correlations beyond thermal ones. It was recently shown~\cite{kurchan,Chan2019} that OTOC are controlled by correlations beyond ETH. 
Despite its promising features, the interpretation of the connection between the OTOC and the underlying quantum state dynamics is, in general, complex. 

The purpose of this Letter is to show that the task of discriminating a pure state that ``looks'' thermal from a true, thermal Gibbs density matrix
might be better achieved by a different physical quantity: the quantum Fisher information (QFI)~\cite{helstrom1969quantum, Tth2014, pezze2014quantum},  a quantity of central importance in metrology~\cite{giovannetti2011advances,Pezz2018} and entanglement theory~\cite{Hyllus2012,Tth2012}. 
The first observation of our work is that the
QFI computed in the eigenstates of the Hamiltonian  ${\cal F}_{\textrm{ETH}}$ (or in the asymptotic state of a quenched dynamics), and the one computed in the Gibbs state at the corresponding inverse temperature $\beta$, ${\cal F}_{\textrm{Gibbs}}$  \cite{Hauke2016, Gabbrielli2018}, satisfy the inequality ${\cal F}_{\textrm{ETH}} \geq {\cal F}_{\textrm{Gibbs}}$, where the equality holds at zero temperature. By computing both terms, we quantify the difference. The corresponding
multipartite entanglement structure, as obtained from the Fisher information densities $f_Q={\cal F}/N$ is in stark contrast. For example, in systems possessing finite temperature phase transitions, we argue that $\mathcal F_{\textrm{ETH}}$ diverges with system size at critical points (implying extensive multipartiteness of entanglement in the pure state), while it is only finite in the corresponding Gibbs ensemble \cite{Hauke2016,Gabbrielli2018,Frrot2018}. \\
The second main result in this work is numerical. The explicit calculation of $\mathcal F_{\textrm{ETH}}$ in a non-integrable model 
is an arduous task as it involves full diagonalization and data processing of off-diagonal matrix elements which exponentially increase with system size. We use state-of-the-art and highly optimized exact diagonalization and data sorting routines 
to extract the universal features of these off-diagonal matrix elements, in order to compute the relevant correlation functions and the corresponding QFI densities. 
We study both ${\cal F}_{\textrm{ETH}}$ and ${\cal F}_{\textrm{Gibbs}}$ in the XXZ model with integrability breaking staggered field, unravelling the interesting behavior of these quantities. \\

\emph {ETH and linear response}.--- 
The ETH ansatz for the matrix elements of observables in the eigenbasis of the Hamiltonian, is formally stated as \cite{Srednicki:1999, Alessio:2016}
\begin{align}
\label{eq:eth}
O_{n m} = O(\bar{E}) \delta_{n m} + e^{-S(\bar{E}) / 2}f_{\hat O}(\bar{E}, \omega)R_{n m},
\end{align}
where $\bar{E}=(E_n+E_m)/2$, $\omega=E_m-E_n$, $S(\bar{E})$ 
is the microcanonical entropy 
and $R_{nm}$
is a random variable with zero average and unit variance.  Both $O(\bar{E})$ and $f_{\hat O}(\bar{E},\omega)$ are smooth functions of their
arguments. In particular, $O(\bar{E})$ is the microcanonical average in a shell centered around energy $\bar{E}$. 
%
Crucially, through the off-diagonal matrix elements, the function $f_{\hat O}(\bar{E},\omega)$ can be extracted, allowing for the explicit calculation of non-equal correlation functions in time. The \emph{response function} and the \emph{symmetrized noise} are defined respectively as  
$\chi_{\hat{O}}(t_1, t_2) \coloneqq -i\theta(t_1- t_2) \langle [\hat{O}(t_1),\hat{O}(t_2)] \rangle$ and $S_{\hat{O}}(t_1, t_2) \coloneqq \langle \{\hat{O}(t_1),\hat{O}(t_2)\}\rangle-2 \langle \hat{O}(t_1)\rangle \langle \hat{O}(t_2)\rangle$. 
The expectation value of these correlation functions can be taken with respect to a single energy eigenstate ${\hat H \ket{E} = E\ket{E}}$ and Fourier-transformed with respect to the time difference to be expressed in frequency domain.
%
For local operators or sums of local operators, the spectral function ${{\rm Im}[\chi_{\hat{O}}(\omega)]= -\chi_{\hat{O}}^{\prime \prime}(\omega)}$ and $S_{\hat{O}}(\omega)$ can be approximated imposing the ETH~\cite{Srednicki:1999,Alessio:2016}. In the thermodynamic limit they read 
\begin{eqnarray}
\label{eq:mule}
\chi_{\hat{O}}^{\prime \prime}(\omega)&\approx&2\pi \sinh \left(\frac{\beta \omega}{2}\right) |f_{\hat{O}}(E,\omega)|^2 \ , \\
S_{\hat{O}}(\omega)&\approx& 4\pi \cosh \left(\frac{\beta \omega}{2} \right) |f_{\hat{O}}(E,\omega)|^2 \ .
\end{eqnarray}
These relations satisfy the fluctuation dissipation theorem (FDT) $S_{\hat{O}}(\omega)=2\coth(\frac{\beta \omega}{2}) \chi_{\hat{O}}^{\prime \prime}(\omega)$. In this 
context, the inverse temperature is 
given by the thermodynamic definition
$\beta = \partial S(E) / \partial E$ and it corresponds to the canonical temperature at the same average energy $E=\bra{E} \hat H \ket{E} = \text{Tr}(\hat H\, e^{-\beta \hat H})/Z$. \\

\emph {Quantum Fisher information and linear response}.--- 
There has been some interest in relating ETH to the bipartite entanglement entropy \cite{Deutsch2013, Vidmar2017,Murthy2019EE}, here
we apply ETH to the quantum Fisher information $\mathcal F(\hat O)$. This quantity was introduced to bound the precision of the estimation of a parameter $\phi$, conjugated to an observable $\hat O$ using a quantum state $\hat \rho$, via the so-called quantum Cramer-Rao bound $\Delta \phi^2 \leq 1/M \mathcal F(\hat O)$, where $M$ is the number of independent measurements made in the protocol \cite{Pezz2018}. 

Most importantly, the QFI has key mathematical properties \cite{Braunstein1994,PETZ2011,Tth2014, Pezz2018}, such as convexity, additivity, monotonicity and it 
can be used to probe the multipartite entanglement structure of a quantum state~\cite{Hyllus2012,Tth2012}.  If, for a certain $\hat O$, the QFI density satisfies
\begin{equation}
 f_Q(\hat O)=\frac{\mathcal{F}(\hat O)}{N} > m  \ ,
\end{equation}
then, at least $(m+1)$ parties in the system are entangled (with $1\leq m \leq N-1$ a divisor of $N$). 
In particular, if $N-1 \leq f_Q(\hat O) \leq N$, then the state is called genuinely $N$-partite entangled. 
In general, different operators $\hat{O}$ lead to different bounds and there is no systematic method (without some knowledge on the physical system \cite{Hauke2016,Pezz2017Gabri}) to choose the optimal one, which will typically be an extensive sum of local operators. 
For a general mixed state described by the density matrix $\rho=\sum_n p_n |E_n\rangle\langle E_n|$,  it was shown that \cite{Braunstein1994}
\begin{eqnarray}
\label{QFI}
{\mathcal F}(\hat O)=2\sum_{n,n^{\prime}} \frac{(p_n-p_{n^{\prime}})^2}{p_n+p_{n^{\prime}}} |\langle E_n| \hat O|E_{n^{\prime}}\rangle|^2 {\leq 4 \, \langle \Delta \hat O^2 \rangle } \ ,
\end{eqnarray}
{with $\langle  \Delta \hat O^2 \rangle =\text{Tr}(\hat \rho\, \hat O^2 ) -  \text{Tr}(\hat \rho\,\hat O )^2$. The equality holds in the case of pure states $\hat \rho = \ket{\psi}\bra{\psi}$.}

Let us now contrast the QFI computed on a thermodynamic ensemble with the one of a single energy eigenstate for an operator satisfying ETH. When computed 
on a canonical 
Gibbs state with  $p_n=e^{-\beta E_n}/Z$ in Eq.~\eqref{QFI}, it was shown that  \cite{Hauke2016}
\begin{align}
\label{eq:f_gibbs}
\begin{split}
{\cal F}_{\textrm{Gibbs}}(\hat{O
})&= \frac{2}{\pi}\int_{-\infty}^{+\infty} d\omega  \, \tanh \left(\frac{\beta \omega}{2}\right)  \chi_{\hat{O}}^{\prime\prime}(\omega)\ .
\end{split}
\end{align}
The same result holds in the microcanonical ensemble \cite{note_CvsMC}.
If in contrast one considers a pure eigenstate at the same temperature, i.e. with energy $E=
\textrm{Tr}(\hat{H}e^{-\beta \hat{H}}/Z)$ compatible with the average energy of a canonical state in the system, the QFI is
\begin{align}
\begin{split}
    \label{eq:qfi_eth}
{\cal F}_{\textrm{ETH}}(\hat{O}) &= 4\, \langle E | \Delta \hat{O}^2 |E \rangle =  \int_{-\infty}^{+\infty} \frac{d\omega}{\pi} S_{\hat{O}}(\omega) \\
& = \frac{2}{\pi}\int_{-\infty}^{+\infty} d\omega  \, \coth \left(\frac{\beta \omega}{2}\right)  \chi_{\hat{O}}^{\prime\prime}(\omega)\ ,
\end{split}
\end{align}
where $S_{\hat{O}}(\omega)$ in the previous equation is determined by the function $f_{\hat{O}}(E, \omega)$ appearing in Eq.~\eqref{eq:mule} as described. Since $S_{\hat{O}}(\omega)$ evaluated explicitly from ETH is  equivalent to its canonical counterpart, then the following result holds 
\begin{equation}
    \label{eq:bound}
    {\cal F}_{\textrm{ETH}}(\hat{O})\ge {\cal F}_{\textrm{Gibbs}}(\hat{O}) \ .
\end{equation}
Notice that the variance over the Gibbs ensemble, that already bounds the corresponding QFI through Eq.(\ref{QFI}), also bounds from above  $\mathcal F_{\text{ETH}}$, as discussed below.

This analysis has immediate consequences for the QFI and the entanglement structure, of asymptotic states in out-of-equilibrium unitary dynamics. 
In this framework, the expectation value of time dependent operators $O(t)=\bra{\psi} \hat O(t) \ket{\psi}$ (or of the correlation functions defined above) are taken with respect to an initial pure state $\ket{\psi}$, which is not an eigenstate of the Hamiltonian $\hat H$. 
Provided that the QFI attains an asymptotic value at long times $\mathcal F_{\infty}$,
taking the long-time average 
\cite{Note_TimeAve}, whenever there are no degeneracies or only a subextensive number of them, 
we have that ${{\overline{\mathcal F(\hat O)} = \mathcal F_{\infty}}(\hat O)}= 4 \langle \Delta \hat O^2 \rangle_{\text {DE}}$  
with ${\langle \,\cdot\, \rangle_{\text{DE}} = \text{Tr}(\hat\rho_{\text{DE}}\,\cdot\,)}$ \cite{Rossini2014, pappalardi2017multipartite}, and the diagonal ensemble defined as $\hat{\rho}_{\text{DE}}=\sum|c_{n}|^2\ket{E_{n}}\bra{E_{n}}$ with $c_n = \bra{\psi}E_n\rangle$. We remark that, since the out-of-equilibrium global state is pure, $\mathcal {F}_{\infty}(\hat{O})$
is given by the variance of $\hat O$ over the diagonal ensemble which is different from the
QFI computed on the state $\hat \rho_{\text{DE}}$ using Eq.~\eqref{QFI}.
 See \cite{SM} for the details on the out-of-equilibrium setting.

For sufficiently chaotic Hamiltonians, the initial state $\ket{\psi}$ considered is usually a microcanonical superposition around an average energy $E  = \bra {\psi} \hat H \ket{\psi}$ with variance ${\delta^2 E = \bra {\psi} \hat H^2 \ket{\psi} - \bra {\psi} \hat H \ket{\psi}}^2$, i.e. $|c_n|^2$ has a narrow distribution around $E$ with small fluctuations ${\delta^2 E / E^2 \sim 1/N}$ \cite{Alessio:2016}. 
Then it follows
$ {\langle\Delta \hat O^2 \rangle_{\text{DE}}=  \bra{E}\Delta \hat O^2 \ket{E}\, +  \left ( \frac{\partial O}{\partial E} \right )^2 \, \delta^2 E }$, where
the first term represents fluctuations inside each eigenstate --computed before in Eq.~\eqref{eq:qfi_eth} -- and the second is related to energy fluctuations \cite{SM}. This observation, together with the bound (\ref{eq:bound}), leads to 
\begin{equation}
    \label{eq:final_bound}
   {\cal F}_{\infty}(\hat{O}) \ge {\cal F}_{\textrm{ETH}}(\hat{O})\ge {\cal F}_{\textrm{Gibbs}}(\hat{O}) \ ,
\end{equation}
where the equality holds in the low temperature limit $T\to 0$. This also implies that $4 \langle \Delta \hat  O^2\rangle_{\text{Gibbs}}  \geq \mathcal F_{\text{ETH}}(\hat{O})$ \cite{note_diag}. 
    These expressions 
    set a hierarchy in the entanglement content of ``thermal states'' at the same temperature,  yet of different nature (mixed/pure). 
Furthermore, via Eqs.~\eqref{eq:f_gibbs}-\eqref{eq:qfi_eth}, one can quantify this difference via $\Delta \mathcal F = \mathcal F_{\text{ETH}}-\mathcal F_{\text{Gibbs}}= 1/\pi \int d\omega S_{\hat{O}}(\omega) / \cosh^2(\beta\omega/2) $.

{\emph {Multipartite entanglement at thermal criticality}}.---
The major difference between the ETH and Gibbs multipartite entanglement
can be appreciated at critical points of thermal phase transitions, where $\hat O$ in (\ref{QFI}) is the order parameter of the theory.
While it is well known that the QFI does not witness divergence of multipartiteness at thermal criticality, i.e. $\mathcal F_{\textrm{Gibbs}}/N \sim \text{const.}$ \cite{Hauke2016,Gabbrielli2018}, on the other hand, the ETH result obeys the following critical scaling with the system size $N$
\begin{equation}
  f_Q^{\textrm{ETH}} \sim \frac{ \mathcal F_{\textrm{ETH}}}N \sim N^{\gamma / (\nu\, d)} \ ,
\end{equation}
where $\gamma $ and $\nu$ are the critical exponents of susceptibility and correlation length of the thermal phase transition respectively and $d$ is the dimensionality of the system \cite{cardy2012finite}. 

\emph {Evaluation}.--- We now turn to the evaluation of Eq.~\eqref{eq:mule} in the context of a physical system with a microscopic Hamiltonian description. Consider the anisotropic spin-$\frac{1}{2}$ Heisenberg chain, also known as the spin-$\frac{1}{2}$ XXZ chain, with the Hamiltonian given by ($\hbar = 1$):
\begin{align}
\label{eq:h_xxz}
\hat{H}_{\textrm{XXZ}} = \sum_{i=1}^{N-1}\left[\left(\hat{\sigma}^x_{i}\hat{\sigma}^x_{i+1} + \hat{\sigma}^y_{i}\hat{\sigma}^y_{i+1}\right) + \Delta\,\hat{\sigma}^z_{i}\hat{\sigma}^z_{i+1}\right],
\end{align} 
where $\hat{\sigma}^\nu_{i}$, $\nu = x,y,z$, correspond to Pauli matrices in the $\nu$ direction at site $i$ in a one-dimensional lattice with $N$ sites defined with open boundary conditions (OBCs). In Eq.~\eqref{eq:h_xxz}, $\Delta$ corresponds to the anisotropy parameter. The spin-$\frac{1}{2}$ XXZ chain corresponds to one of the canonical integrable models. We now add a strong integrability breaking perturbation in the form of a staggered magnetic field across the chain, with the Hamiltonian defined as
\begin{align}
\label{eq:h_sf}
\hat{H}_{\textrm{SF}} = \hat{H}_{\textrm{XXZ}} + b\, \sum_{i\, \textrm{even}} \hat{\sigma}^z_i,
\end{align} 
where $b$ is the strength of the staggered magnetic field. Eq.~\eqref{eq:h_sf} is the Hamiltonian of the {\em staggered field} model. This model is quantum chaotic with Wigner-Dyson level spacing statistics and diffusive transport~\cite{marlon}. The models described before commute with the total magnetization operator in the $z$ direction, $[\hat{H}_{\textrm{XXZ}},  \sum_i\hat{\sigma}^z_i]=[\hat{H}_{\textrm{SF}}, \sum_i\hat{\sigma}^z_i]=0$ and are, therefore, $U(1)$-symmetric. Even with OBCs, parity symmetry is present in the system. We break this symmetry by adding a small perturbation $\delta\hat{\sigma}^z_1$ on the first site. 
\begin{figure}[t]
\fontsize{13}{10}\selectfont 
\centering
\includegraphics[width=\columnwidth]{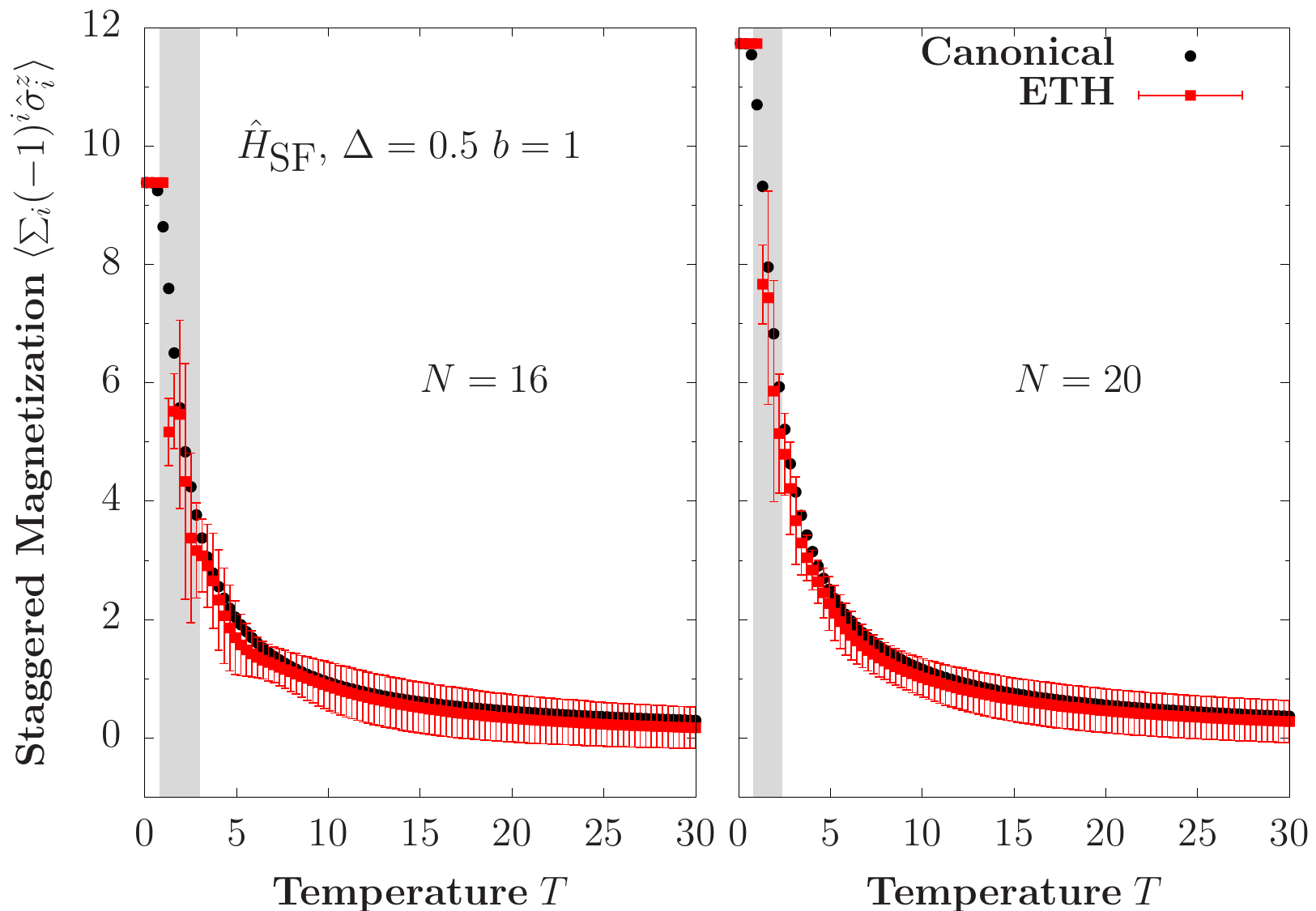}
\caption{Expectation value of the staggered magnetization as a function of temperature in both the canonical ensemble and the corresponding ETH prediction for and $N = 16$ (\textcolor{blue}{left}) $N = 20$ (\textcolor{blue}{right}). Gray area highlights the low temperature regime, close to the edges of the spectrum where the ETH prediction gives the largest fluctuations.}
\label{fig:StagDiag}
\end{figure}
To evaluate our results in the canonical ensemble and in the context of ETH, we proceed with the full diagonalization of $\hat{H}_{\textrm{SF}}$ in the largest $U(1)$ sector, in which $\sum_i \langle \hat{\sigma}^z_i \rangle = 0$. We focus on the total staggered magnetization $\hat{O} = \sum_i (-1)^i \hat{\sigma}^z_i$ as our extensive observable, and compute all the matrix elements of $\hat{O}$ in the eigenbasis of the Hamiltonian $\hat{H}_{\textrm{SF}}$ (see \cite{SM} for an evaluation on a local, non-extensive observable). 

Our starting point is to evaluate the expectation value of $\hat{O}$ in the canonical ensemble and compare it with the ETH prediction. In the thermodynamic limit, a single eigenstate $\ket{E}$ with energy $E$ suffices to obtain the canonical prediction: $\langle \hat{O} \rangle =\bra{E} \hat{O} \ket{E} {=} \text{Tr}(\hat {O}\, e^{-\beta \hat H})/Z$, with an inverse temperature $\beta$ that yields an average energy $E$. For finite-size systems, we instead focus on a small energy window centered around $E$ of width $0.1\epsilon$ in order to average eigenstate fluctuations, where $\epsilon$ is the bandwidth of the Hamiltonian for a given $N$. Fig.~\ref{fig:StagDiag} shows $\langle \hat{O} \rangle$ as a function of temperature for two different system sizes, including $N=20$, the largest system we have access to (Hilbert space dimension $\mathcal{D} = N!/[(N/2)!(N/2)!] = 184\;756$). The results exhibit the expected behavior predicted from ETH for finite-size systems: the thermal expectation value is well approximated away from the edges of the spectrum (low temperature, section highlighted in gray on Fig.~\ref{fig:StagDiag}), and, moreover, the canonical expectation value is better approximated as the system size increases. \\
\begin{figure}[t]
\fontsize{13}{10}\selectfont 
\centering
\includegraphics[width=0.82\columnwidth]{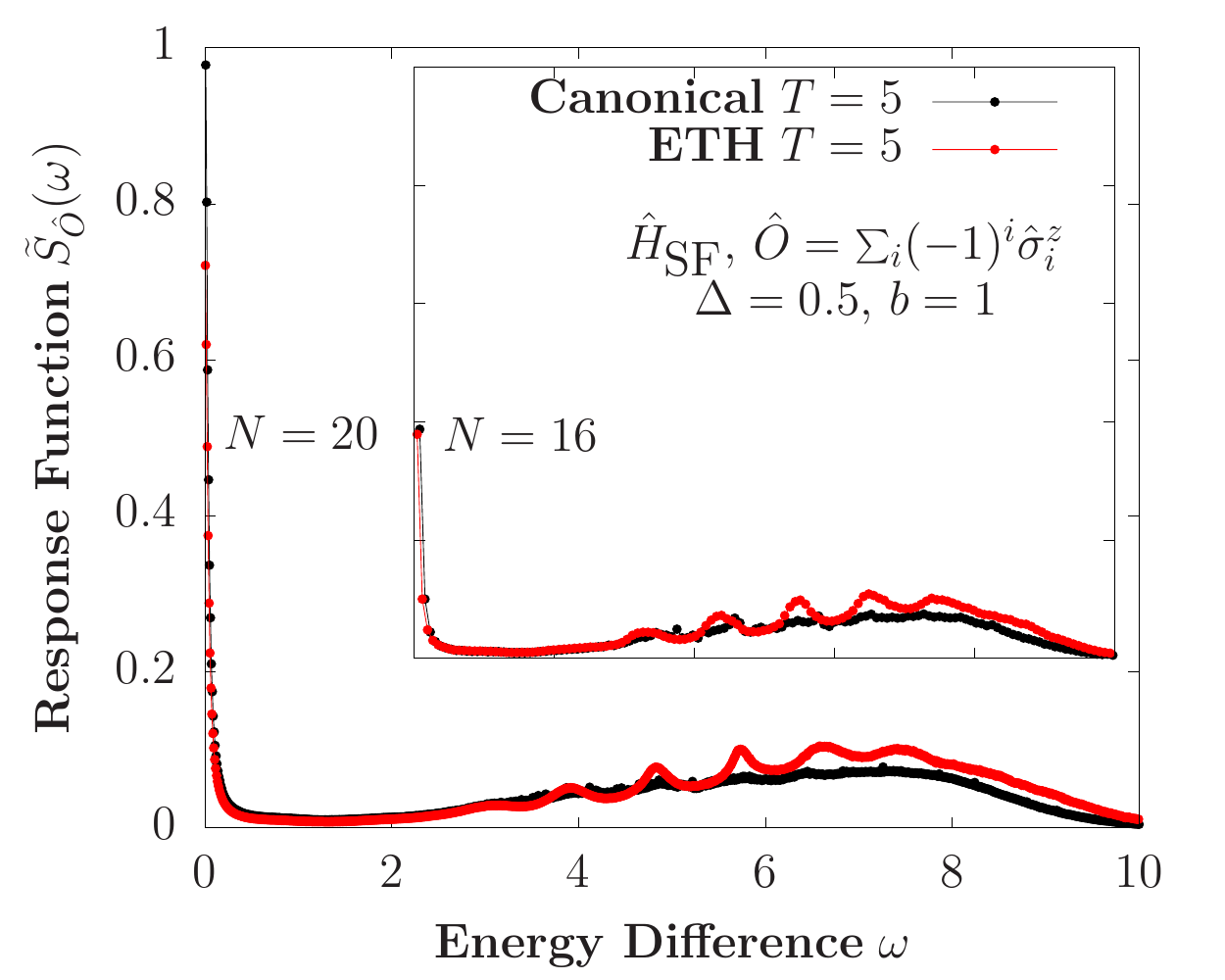}
\caption{Response function $S_{\hat{O}}(\omega)$ computed directly from ETH and in the canonical ensemble for $N=16$ (inset) and $N=20$ (main) for $T=5$.}
\label{fig:RespFuncExt}
\end{figure}
We now turn to the evaluation of $\mathcal{F}_{\textrm{ETH}}$ and $\mathcal{F}_{\textrm{Gibbs}}$. The task requires to either compute $S_{\hat{O}}(E, \omega)$ or $\chi^{\prime \prime}_{\hat{O}}(E, \omega)$ in each respective framework. For the former, in the context of ETH, we can employ Eq.~\eqref{eq:mule} which depends only on $f_{\hat{O}}(E, \omega)$. As before, we focus on a small window of energies and extract all the relevant off-diagonal elements of $\hat{O}$ in the eigenbasis of $\hat{H}_{\textrm{SF}}$. Fluctuations are then accounted for by computing a bin average over small windows $\delta \omega$, 
chosen such that the resulting average produces a smooth curve 
[see \cite{SM} for a detailed description on the extraction of $e^{-S(E)/2}f_{\hat{O}}(E, \omega)$] \cite{Mondaini2017,Khatami2013}. The procedure leads to a smooth function $e^{-S(E)/2}f_{\hat{O}}(E, \omega)$, in which the first factor is a constant value with respect to $\omega$. The entropy factor can be left undetermined in our calculations if we normalize the curve by the sum rule shown in Eq.~\eqref{eq:qfi_eth}, computed in this case from the ETH prediction of the expectation value of $\langle \Delta\hat{O}^2 \rangle$. In the context of the canonical ensemble, $S_{\hat{O}}(\omega)$ can be explicitly evaluated by computing the thermal expectation value of the non-equal correlation function in the frequency domain \cite{SM}.

In Fig.~\ref{fig:RespFuncExt} we show $S_{\hat{O}}(\omega)$ for both the canonical ensemble for $T = 5$ and the corresponding ETH prediction normalized by the sum rule mentioned before. The sum rule is evaluated from the expectation values computed within both the canonical ensemble and ETH, correspondingly. It can be observed that the main features of the response function can be well approximated from the corresponding ETH calculation. For this particular case, however, the approximation is only marginally improved by increasing the system size. This behavior is expected given that overall fluctuations for extensive observables carry an extensive energy fluctuation contribution, as mentioned before \cite{Alessio:2016}.
\begin{figure}[b]
\fontsize{13}{10}\selectfont 
\centering
\includegraphics[width=0.85\columnwidth]{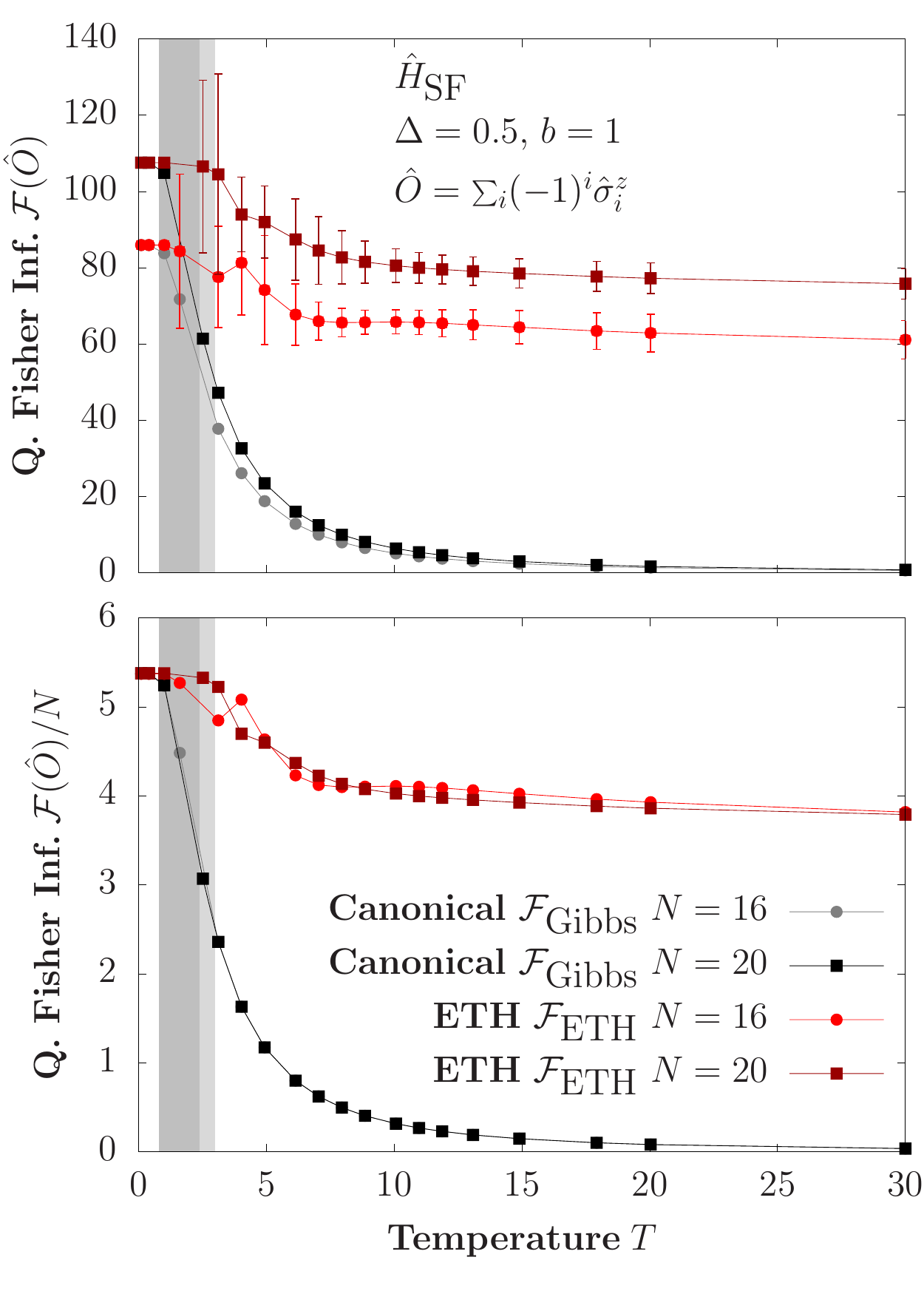}
\caption{The quantum Fisher information and the corresponding density for different system sizes as a function of temperature in both the canonical ensemble ($\mathcal{F}_{\textrm{Gibbs}}$) and corresponding ETH prediction ($\mathcal{F}_{\textrm{ETH}}$). At infinite temperature ETH predicts the presence of multipartite entanglement while there is none in the canonical ensemble.}
\label{fig:FisherExt}
\end{figure}
The previous analysis unravels the agreement between the thermal expectation values of non-equal correlation functions in time and those predicted by ETH. From these results, as $S_{\hat{O}}(\omega)$ (and, consequently, $\chi^{\prime \prime}_{\hat{O}}(\omega)$ from the FDT) is well approximated by means of ETH, the inequality in Eq.~\eqref{eq:bound} is satisfied. 

Finally, we compute the QFI for $\hat{O}$ in our model within both contexts: $\mathcal{F}_{\textrm{ETH}}$ and $\mathcal{F}_{\textrm{Gibbs}}$. The results are shown in Fig.~\ref{fig:FisherExt}. The fluctuations in the ETH calculation of $\mathcal{F}_{\textrm{ETH}}$ are inherited from the fluctuations of the predicted expectation value of $\langle \Delta\hat{O}^2 \rangle$, which, as expected for finite-size systems, decrease away from the edges of the spectrum. Both predictions for the QFI, canonical and ETH, are equivalent at vanishing temperatures. Remarkably, the QFI predicted from ETH is finite at infinite temperature, while the QFI from the canonical ensemble in this regime vanishes. We emphasize that although the QFI can be used in order to infer the structure of multipartite entanglement i.e. the number of sub-systems entangled, it is not a measure of these correlations (in the mathematical sense of the formal theory of entanglement~\cite{family}).

\emph {Conclusions.---}
We have shown that the QFI detects the difference between a pure state satisfying ETH and the Gibbs ensemble at the corresponding temperature. The extension of these results to integrable systems,
described by the generalized Gibbs ensemble, is the subject of current work.
Even though 
it is expected that global observables could be sensitive to the difference between pure states and the Gibbs ensemble \cite{Garrison2018}, 
several operators including sum of local ones and the non-local entanglement entropy appear to coincide at the leading order with the thermodynamic values when ETH is applied \cite{Garrison2018, Polkovnikov2011, Santos2011, Gurarie2013,  Alba2017}.
In this work, the difference between ETH/Gibbs multipartite entanglement, which can be macroscopic in proximity of a thermal phase transition, is observed numerically in a XXZ chain with integrability breaking term, when the temperature grows toward infinity. The consequences of this could be observed in ion trap and cold-atom experiments via phase estimation protocols on pure state preparations evolved beyond the coherence time \cite{SM}. Our result suggests that although at a local level all thermal states look the same, a quantum information perspective indicates that there are many ways to be thermal. 

\textit{Acknowledgments.---} We are grateful to R. Fazio, M. Fabrizio, G. Guarnieri, M. T. Mitchison, L. Pezz\`e,  A. Purkayastha, M. Rigol and A. Smerzi for useful discussions. We thank P. Calabrese, C. Murthy and A. Polkovnikov for the useful comments on the manuscript. M.B. and J.G. acknowledge the DJEI/DES/SFI/HEA Irish Centre for High-End Computing (ICHEC) for the provision of computational facilities and support, project TCPHY104B, and the Trinity Centre for High-Performance Computing. This work was supported by a SFI-Royal Society University Research Fellowship (J.G.) and the Royal Society (M.B.). We acknowledge funding from European Research Council Starting Grant ODYSSEY (Grant Agreement No. 758403).
\bibliography{ethfisher}

\begin{thebibliography}{56}
\bibitem [{not()}]{note_CvsMC}%
  \BibitemOpen
  \href@noop {} {}\bibinfo {note} {Since the definition (\ref{QFI}) applies to
  both the microcanonical (MC) and the canonical Gibbs ensembles, if the energy
  width of the two coincide, then the QFI's are the same and given exactly by
  Eq.(\ref{eq:f_gibbs}). See the \cite{SM} for a derivation based on the ETH.
  From now on we assume equivalence of ensembles, otherwise only the
  microcanonical result should be considered.}\BibitemShut {Stop}%
\bibitem [{Not()}]{Note_TimeAve}%
  \BibitemOpen
  \href@noop {} {}\bibinfo {note} {The long-time average of a time-dependent
  function $f(t)$ is defined as $\overline{f} = \lim_{T\to \infty} T^{-1}
  \int_0^T \, f(t) $. In the case of two-times functions $f(t_1, t_2)$, than
  the time-average is performed over the average time $T=(t_1+t_2
  )/2$.}\BibitemShut {Stop}%
\bibitem [{not()}]{note_diag}%
  \BibitemOpen
  \href@noop {} {}\bibinfo {note} {Since in standard thermodynamics energy
  fluctuations are small $\delta E_{\text{Gibbs}}^2 / E_{\text{Gibbs}}^2 \sim
  1/N$, the same expansion around the average energy $E=\text{Tr} (\hat
  \rho_{\text{Gibbs}}\hat O)$ done for the diagonal ensemble can be performed
  for $\langle \Delta \hat O^2\rangle_{\text{Gibbs}}= \text{Tr}(\hat
  \rho_{\text{Gibbs}} \hat O^2) - \text{Tr}(\hat \rho_{\text{Gibbs}} \hat
  O)^2$, therefore $\langle \Delta \hat O^2\rangle_{\text{Gibbs}} \geq \langle
  E | \Delta \hat O^2 |E\rangle $.}\BibitemShut {Stop}%
\end{thebibliography}

\end{document}


\title{Supplementary Material \\ 
Multipartite entanglement structure in the Eigenstate Thermalization Hypothesis }

\author{Marlon Brenes}
\email{brenesnm@tcd.ie}
\affiliation{Department of Physics, Trinity College Dublin, Dublin 2, Ireland}

\author{Silvia Pappalardi}
\email{spappala@sissa.it}
\affiliation{SISSA, Via Bonomea 265, I-34135 Trieste, Italy}
\affiliation{ICTP, Strada Costiera 11, I-34151 Trieste, Italy}

\author{John Goold}
\affiliation{Department of Physics, Trinity College Dublin, Dublin 2, Ireland}

\author{Alessandro Silva}
\affiliation{SISSA, Via Bonomea 265, I-34135 Trieste, Italy}
  
\date{\today}

\maketitle

In this Supplementary Material, we first provide additional information on the numerical computations of $\mathcal F_{\text{ETH}}$.  In Sec.~\ref{sec:f_o} we report the details on the extraction of the ETH smooth functions from the matrix elements in the energy eigenbasis. In Sec.~\ref{sec:local}, we show the corresponding results for a local operator. In Sec.~\ref{sec:review}, we review some of the known results -- relevant for our work -- regarding the asymptotic quantum Fisher information (QFI) after quenched dynamics and the validity of fluctuation-dissipation theorem within ETH. In Sec.~\ref{sec:QFI_ETH} we present an alternative derivation of the QFI in a thermodynamic ensemble, based solely on the ETH. In Sec.\ref{sec:experiment}, we discuss a possible scheme to observe the consequences of our result in atomic experiments.
 
\section{Extraction of $e^{-S(\bar{E})/2}f_{\hat{O}}(\bar{E}, \omega)$ from the off-diagonal matrix elements of observables}
\label{sec:f_o}

\begin{figure}[b]
\fontsize{13}{10}\selectfont 
\centering
\includegraphics[width=0.9\columnwidth]{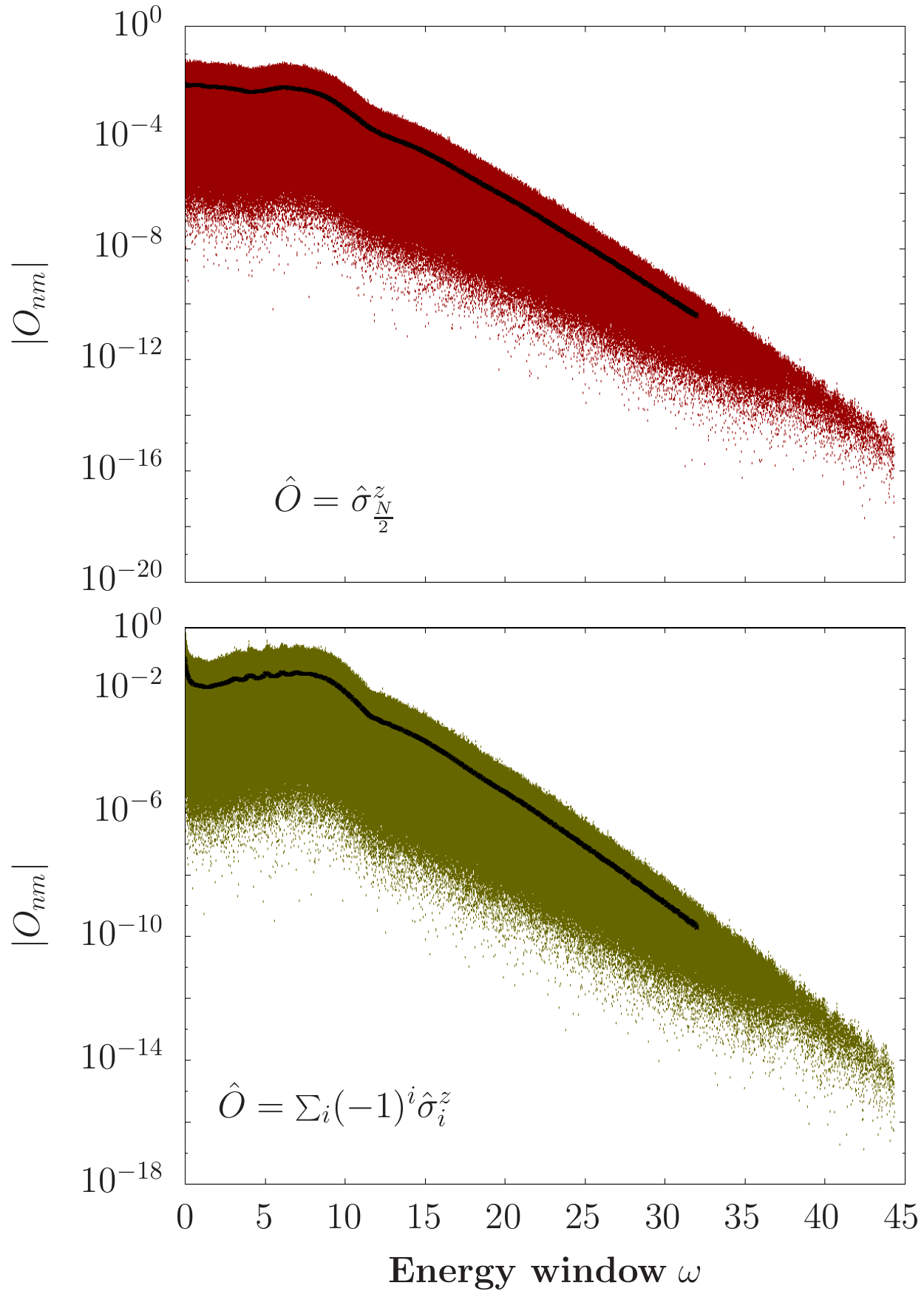}
\caption{Absolute value of the off-diagonal elements in the energy eigenbasis of the local magnetization in the middle of the chain (top) and the total staggered magnetization (bottom) as a function of $\omega$ for $T = 5$ and $N = 18$. The black lines correspond to binned averages. }
\label{fig:f_o}
\end{figure}

In this section, we study the off-diagonal elements of local observables in the energy eigenbasis as a function of $\omega = E_m - E_n$, where $E_k$ labels the $k$-th energy eigenvalue of the Hamiltonian. The appropriate analysis of these elements leads to the smooth function $e^{-S(\bar{E})/2}f_{\hat{O}}(\bar{E},\omega)$, from which the non-equal correlation functions in time and, in turn, the quantum Fisher information (QFI) depend on according to the ETH prediction.

Our starting point is to select a target energy $E$, such that $E = \braket{E | \hat{H} | E} = \text{Tr}(\hat {H}\, e^{-\beta \hat H})/Z$, where $\hat{H}$ is the Hamiltonian of the staggered field model from Eq.~(12)(main text). In the thermodynamic limit, a single eigenstate $\ket{E}$ and its corresponding off-diagonal overlaps with $\hat{O}$ suffice to compute the correlation functions according to the ETH prediction. For finite-size systems, however, we focus on a small window of energies centered around the target energy $E$ of width $0.1\epsilon$, where $\epsilon = E_{\textrm{max}} - E_{\textrm{min}}$ corresponds to the bandwidth of the Hamiltonian at a given system size. Presumably, all the eigenstates in this energy window contain approximately the same average energy. 

To extract $e^{-S(\bar{E})/2}f_{\hat{O}}(\bar{E},\omega)$, we compute the binned average of the samples. The binned average is computed using small frequency windows $\delta \omega$. The size of these windows is selected such that a smooth curve is obtained from the average and the resulting function is not sensitive to the particular choice of $\delta \omega$. This window of frequencies typically changes depending on the dimension of the magnetization subsector studied in our spin model \cite{Mondaini2017}.

In Fig.~\ref{fig:f_o} we present the absolute value of the off-diagonal elements of both the local magnetization operator in the middle of the chain and the total staggered magnetization. These matrix elements were computed for $T = 5$, $N = 18$ and an energy window of width $0.1\epsilon$. The smooth black lines shown are binned averages for each corresponding observable. This average corresponds to $e^{-S(\bar{E})/2}f_{\hat{O}}(\bar{E},\omega)$ up to a constant factor that can, in principle, be determined from finite-size scaling. Both this constant factor, however, as well as the entropy term, can be left undetermined in our calculations as they only affect the approximations on correlation functions on only constant values of $\omega$. As explained in the main text and the next section, these correlation functions are defined under physical normalization conditions, allowing us to focus on the main $\omega$ dependence of $f_{\hat{O}}(\bar{E},\omega)$.

The binned average of the local observables from Fig.~\ref{fig:f_o} exhibits an interesting exponential decay behavior at high frequencies, which has been observed in previous works \cite{Mondaini2017, Alessio:2016,Moessner2015} and 
is related to the universal exponential decay of two-point correlation functions in time for chaotic systems with a bounded spectrum \cite{Huse2006,murthy2019}. On the opposite side of the spectrum, at low frequencies, $f_{\hat{O}}(\bar{E},\omega)$ contains important features relevant to the long-time behavior of correlation functions. These frequencies are the most relevant for the response functions used in this work to evaluate the quantum Fisher information in the context of ETH.

\section{Further evaluation using a local, non-extensive observable}
\label{sec:local}

\begin{figure}[t]
\fontsize{13}{10}\selectfont 
\centering
\includegraphics[width=0.9\columnwidth]{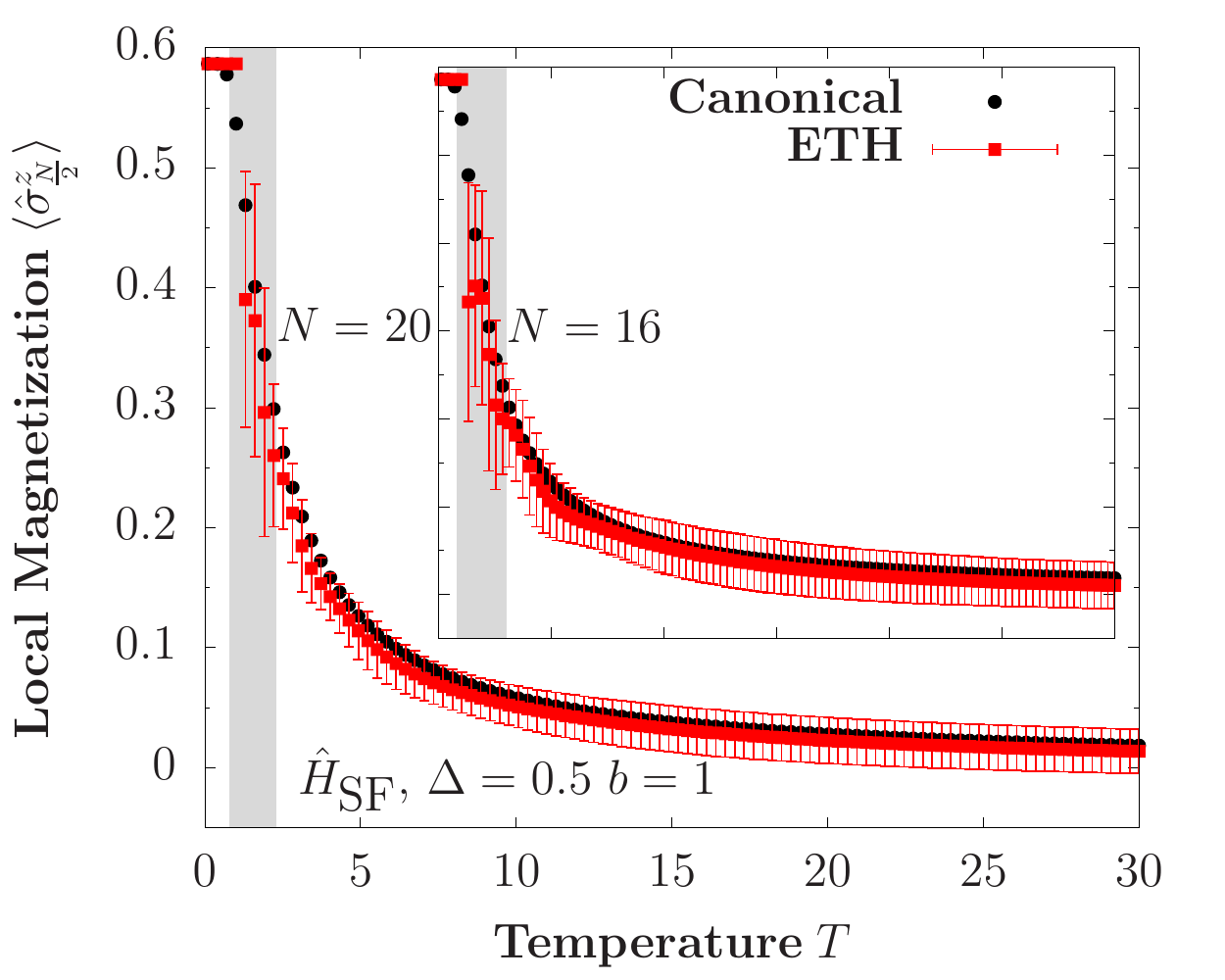}
\caption{Expectation value of the local magnetization in the middle of the chain as a function of temperature in both the canonical ensemble and the corresponding ETH prediction for and $N = 16$ (inset) $N = 20$ (main).}
\label{fig:LocalDiag}
\end{figure}

In the main text we focused on the staggered magnetization, an extensive observable. In this section we focus on the local magnetization in the middle of the chain
\begin{align}
\hat{O} = \hat{\sigma}^z_{\frac{N}{2}}.
\end{align}
As before, our starting point is to evaluate the expectation value of $\hat{O}$ in the canonical ensemble and compare it with the ETH prediction. The results are shown in Fig.~\ref{fig:LocalDiag}. The expectation value in the ETH framework is computed from the average of a small window of energies centred around $E$ of width $0.1\epsilon$, where $\epsilon$ is the bandwidth of the Hamiltonian for a given system size $N$. As in the main text, $E$ is an average energy resulting from the expectation value of energy in the canonical ensemble with inverse temperature $\beta$, i.e, $E = \text{Tr}(\hat {H}\, e^{-\beta \hat H})/Z$, where $\hat{H}$ is the Hamiltonian of the staggered field model from Eq.~(12)(main text).
As predicted by ETH for finite-size systems, the approximation is reliable away from the edges of the spectrum (low $T$, section highlighted in gray in Fig.~\ref{fig:LocalDiag}) and fluctuations decrease as the system size is increased.
\begin{figure}[b]
\fontsize{13}{10}\selectfont 
\centering
\includegraphics[width=0.9\columnwidth]{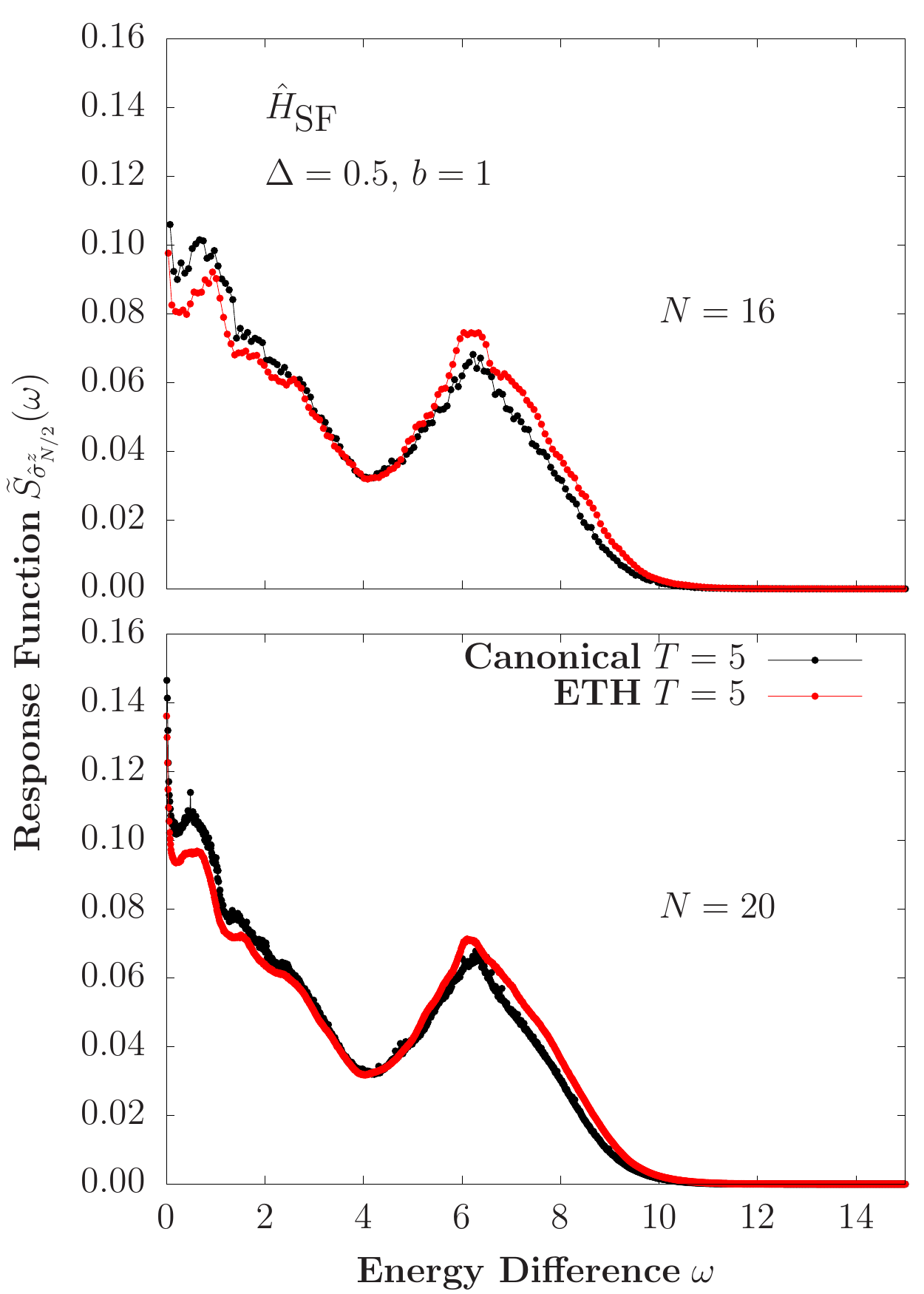}
\caption{Response function $S_{\hat{O}}(\omega)$ for the single site magnetization evaluated at at temperature $T=5$ for $N=16$ (top) and $N=20$ (bottom).} 
\label{fig:RespFuncLocal}
\end{figure}

We can now compute the response functions for $\hat{O}$. In the context of the canonical ensemble, the symmetric response function can be directly evaluated from 
\begin{align}
\label{eq:s_canonical}
S_{\hat{O}}(\omega) &= 2\pi \coth{\left( \frac{\beta \omega}{2} \right)} \nonumber \\
&\times \sum_{n, n^{\prime}} (p_n - p_{n^{\prime}}) |\braket{n | \hat{O} | n^{\prime}}| \delta(\omega + E_{n} - E_{n^{\prime}})\ ,
\end{align}
where $p_n = e^{-\beta E_n}/Z$ are the Boltzmann weights and $\braket{n | \hat{O} | n^{\prime}}$ are the matrix elements of $\hat{O}$ in the energy eigenbasis. We contrast this result with the corresponding prediction obtained from ETH, given by
\begin{align}
S_{\hat{O}}(\bar{E}, \omega)&\approx 4\pi \cosh \left(\frac{\beta \omega}{2} \right) |f_{\hat{O}}(\bar{E},\omega)|^2,
\end{align}
where $f_{\hat{O}}(\bar{E},\omega)$ is extracted up to a constant entropy factor from the procedure described before. We can leave this factor undetermined in our calculations and use the sum rule to normalize the response function. For the local magnetization in the middle of the chain, the sum rule evaluates to
\begin{align}
\int_{-\infty}^{\infty} d\omega\; S_{\hat{O}}(\omega) = 4\pi \langle \Delta\hat{O}^2 \rangle = 4\pi \left( 1 - \langle \hat{\sigma}^z_{N/2} \rangle^2 \right) \ .
\end{align}

\begin{figure}[b]
\fontsize{13}{10}\selectfont 
\centering
\includegraphics[width=0.9\columnwidth]{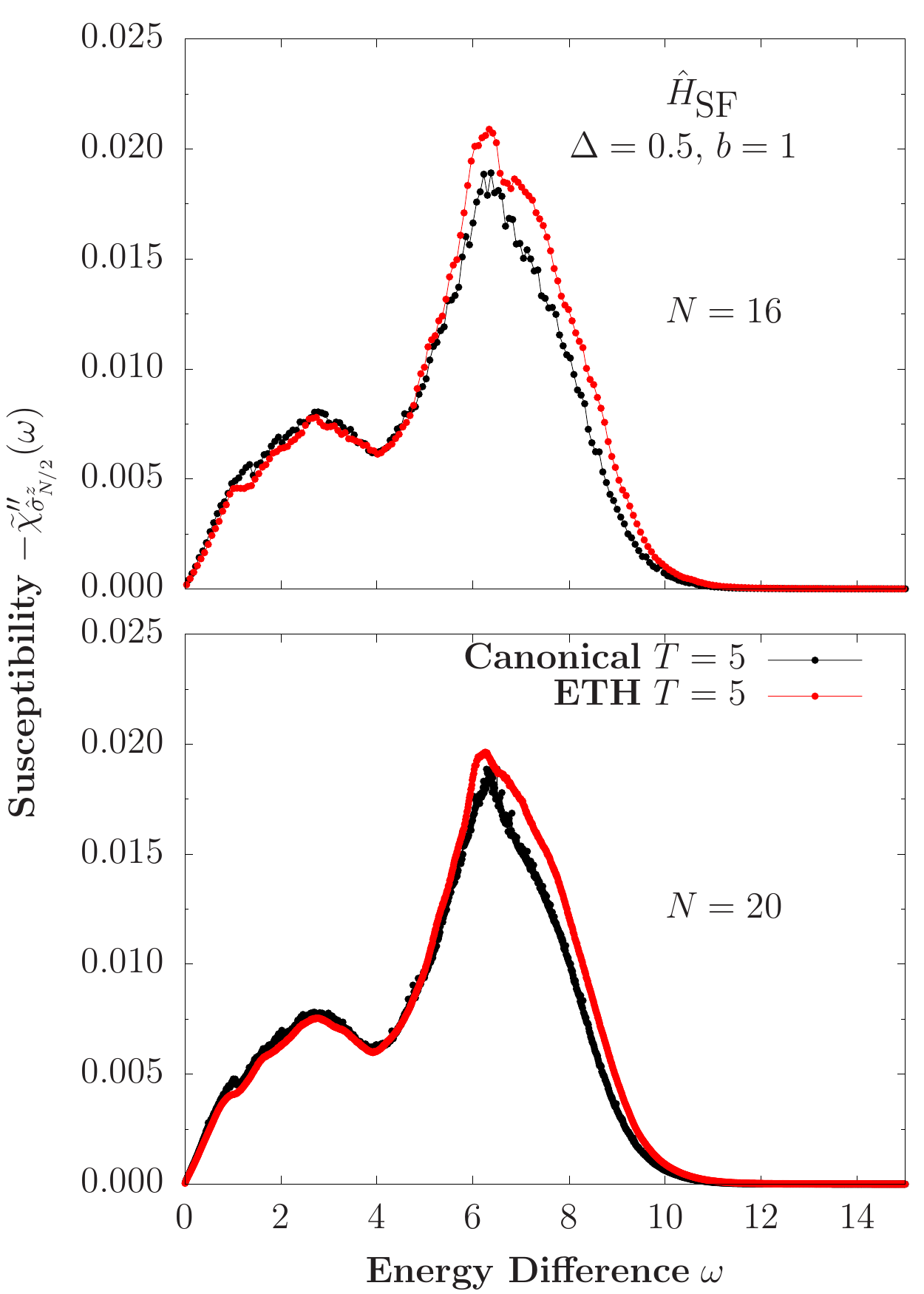}
\caption{Susceptibility $\chi^{\prime \prime}_{\hat{O}}(\omega)$ for the single site magnetization evaluated at at temperature $T=5$ for $N=16$ (top) and $N=20$ (bottom).}
\label{fig:SuscLocal}
\end{figure}

In Fig.~\ref{fig:RespFuncLocal} we show the response function in the context of the canonical ensemble and the corresponding ETH prediction for $N=16$ and $N=20$ at a temperature $T = 5$. The curves are normalized by the sum rule described before, in which the expectation value is taken in the corresponding framework. Unlike extensive observables, energy fluctuations in local intensive observables are subleading and only eigenstate fluctuations are relevant. Since eigenstate fluctuations decrease exponentially with system size, the ETH prediction better approximates the canonical response function as the system size is increased.
 
The susceptibility can be readily computed from the response function invoking the fluctuation-dissipation theorem
\begin{align}
\chi_{\hat{O}}^{\prime \prime}(\omega) = \frac{1}{2} \tanh{\left( \frac{\beta \omega}{2} \right)} S_{\hat{O}}(\omega)\ .
\end{align}
We show the results in Fig.~\ref{fig:SuscLocal}, in which the same features described before for the response function can be observed.
 \begin{figure}[t]
\fontsize{13}{10}\selectfont 
\centering
\includegraphics[width=0.9\columnwidth]{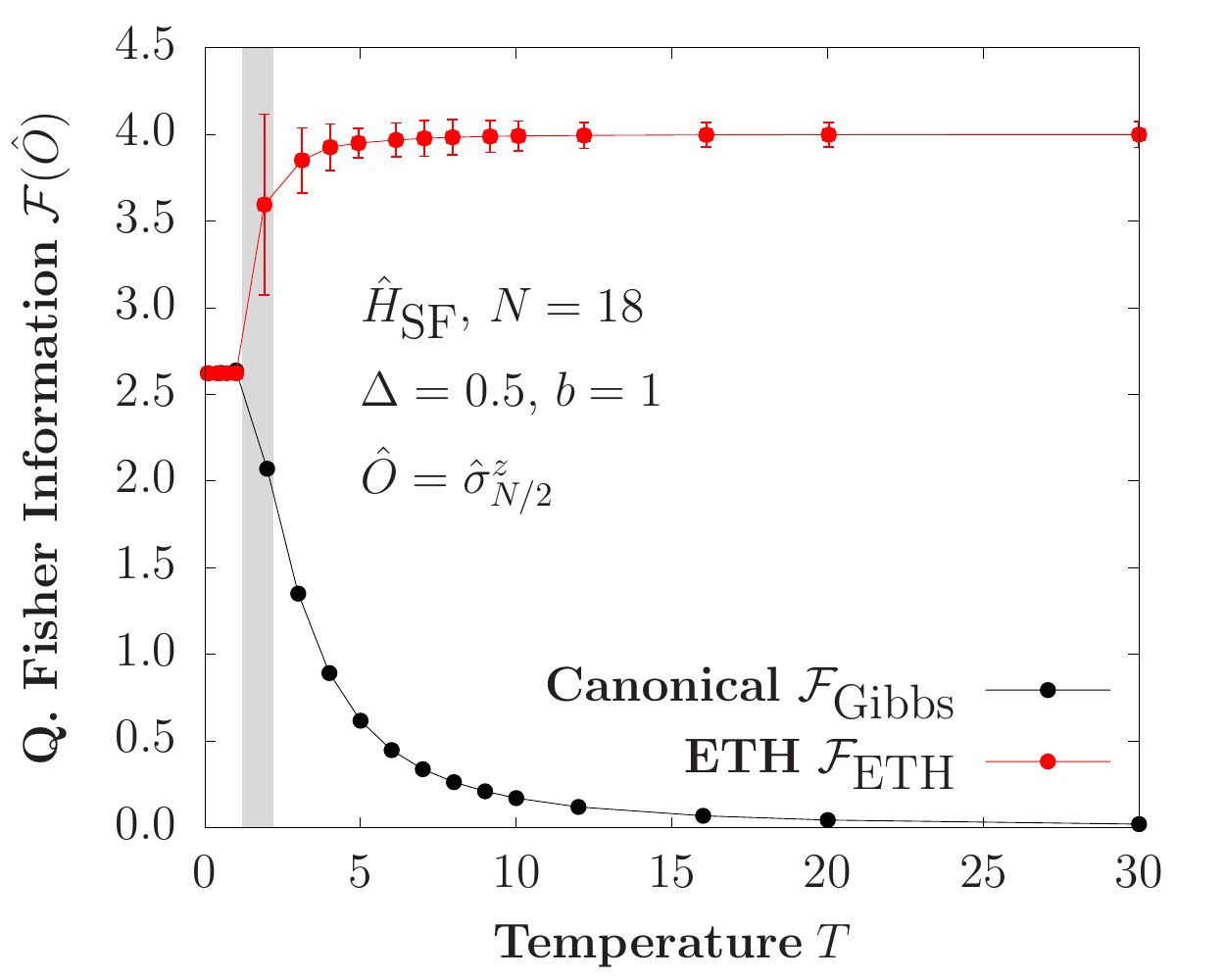}
\caption{The quantum Fisher information for $N=18$ as a function of temperature in both the canonical ensemble ($\mathcal{F}_{\textrm{Gibbs}}$) and corresponding ETH prediction ($\mathcal{F}_{\textrm{ETH}}$) for the local magnetization in the middle of the chain. 
}
\label{fig:FisherLocal}
\end{figure}

Now that we have shown the agreement between the ETH and the canonical ensemble predictions, we can compute the corresponding quantum Fisher information in each framework. As in the main text, within the canonical ensemble, $\mathcal{F}_{\textrm{Gibbs}}$ is given by Eq.~(5)(main text) while the corresponding ETH prediction $\mathcal{F}_{\textrm{ETH}}$ is expressed in Eq.~(7)(main text). The results are shown in Fig.~\ref{fig:FisherLocal}. The fluctuations in the ETH prediction are inherited from the fluctuations on the expectation value of $\langle \Delta\hat{O}^2 \rangle$, which decrease away from low temperatures. In Fig.~\ref{fig:FisherLocal}, it can be observed that $\mathcal{F}_{\textrm{ETH}}$ bounds from above the canonical prediction. For this particular case, the QFI is not a witness of multipartite entanglement since the quantity was computed on a local non-extensive observable. These results, however, show the expected behavior predicted in Eq.~(8)(main text).
 
\section{Asymptotic QFI, ETH and Fluctuation Dissipation Theorem}
\label{sec:review}
This section is devoted to a review of some known results concerning the stationary properties of correlation functions, diagonal ensemble and eigenstates fluctuation's theorems.
 We will first consider the approach to the asymptotic value of time dependent response functions and time-averaged multipartite entanglement \cite{Rossini2014,pappalardi2017multipartite} and discuss the relation between diagonal ensemble's fluctuations and single energy eigenstates \cite{Alessio:2016}.
 We will then review the proof of the fluctuation-dissipation theorem (FDT) for single energy eigenstates \cite{Alessio:2016} and extend it to the diagonal ensemble.
 
 \subsection{Asymptotic multipartite entanglement and single-eigenstates fluctuations}
 Let us consider the standard out-of-equilibrium quench protocol, where a pure quantum state $\ket{\psi}$ -- which is not an eigenstate of the Hamiltonian $\hat H$ -- undergoes unitary dynamics $\ket{\psi(t)} = e^{-i \hat H t}$ $\ket{\psi}$.
The linear response function and the symmetrized noise, also defined in the main text, are given by
\begin{subequations}
\begin{align}
\label{eq:chi_def}
\chi_{\hat O}(t_1, t_2) & = - i \theta(t_1-t_2)\, \langle [\hat O(t_1), \hat O(t_2)]\rangle \ ,\\
\label{eq:S_def}
S_{\hat O}(t_1, t_2) & = 
\langle \{\hat O(t_1), \hat O(t_2)\}\rangle
 - 2 
 \langle \hat O(t_1) \rangle \, \langle \hat O(t_2) \rangle  \ .
\end{align}
\end{subequations}
%
In the out-of-equilibrium setting these quantities do not, in general, depend on the time-difference $t_1-t_2$. However, one may still effectively analyze the dynamics with the new variables $t_{1,2} = T \pm \tau/2$ (also know as Wigner coordinates) and Fourier transform with respect to $\tau$,
restricted to $|\tau|\leq 2 T$. In particular, 
one can obtain information about the stationary state attained at long times. 
Let us first consider the two-point correlation function, i.e. $ \langle \hat O(t_1) \hat O(t_2)\rangle$, since Eqs.~\eqref{eq:chi_def}-\eqref{eq:S_def} depend on such terms. We start by expressing the correlation function in the energy eigenbasis and substituting the Wigner coordinates $t_{1,2} = T \pm \tau/2$. By averaging over $T$, i.e., $\overline{[\cdot]} = \lim_{T\to \infty} T^{-1} \int_0^T [\cdot]$, the correlation function reads
\begin{align} 
\begin{split}
\overline{\langle \hat O(T+\tau/2) \hat O(T-\tau/2)\rangle} & = 
    \sum_{nm}\, |c_n|^2 \, e^{i (E_n - E_m) \tau}\, |O_{nm}|^2  \\
    & = \langle\hat O(\tau) \hat O(0) \rangle_{\textrm{DE}}  \ ,
\end{split}
\end{align}
where $\langle \,\cdot\, \rangle_{\text{DE}} = \text{Tr}(\rho_{\text{DE}}\,\cdot\,)$ and
$\rho_{\textrm{DE}}$ is the so-called \emph{diagonal ensemble} already defined in the main text
\begin{equation}
\label{eq:DE}
\hat \rho_{\textrm{DE}} = \sum_n |c_n|^2 \ket{E_n} \bra{E_n} \quad \quad \text{with}\quad c_n = \bra{\psi}E_n\rangle .
\end{equation}
The same holds for the asymptotic values of the correlation functions (\ref{eq:chi_def}-\ref{eq:S_def}) \cite{Rossini2014}
\begin{subequations}
\begin{align}
\label{eq:corre_fun_DE_chi}
\overline{\chi_{\hat O}(T, \tau)} & = \chi^{\text{DE}}_{\hat O}(\tau) = - i \theta(\tau)\, \langle [\hat O(\tau), \hat O(0)]\rangle_{\textrm{DE}} \ ,\\
\begin{split}
\label{eq:corre_fun_DE_S}
\overline{S_{\hat O}(T, \tau)} & = S^{\text{DE}}_{\hat O}(\tau)  =
\langle \{\hat O(\tau), \hat O(0)\}\rangle_{\textrm{DE}} \\
& \quad \quad \quad \quad \quad - 
2 
\langle \hat O(\tau) \rangle_{\textrm{DE}} \, \langle \hat O(0) \rangle_{\textrm{DE}}  .
\end{split}
\end{align}
\end{subequations}
Provided that the QFI attains an asymptotic value at long times,  then the above discussion immediately implies that
\begin{align}
\label{eq:asy_FQ}
    \begin{split}
    \overline{\mathcal F(\hat O)} & = 4\, \overline{\langle \psi(t) | \Delta^2 \hat{O} |\psi(t)  \rangle} 
     =  \int_{-\infty}^{+\infty} \frac{d\omega}{\pi} S_{\text{DE}}(\omega) 
    = \mathcal{F}_{\infty}(\hat O) \ ,
    \end{split}
\end{align}
where $S_{\text{DE}}(\omega)$ is the Fourier transform with respect to $\tau$ in Eq.~\eqref{eq:corre_fun_DE_S} \cite{pappalardi2017multipartite}.

The discussion so far has been completely general. We now ask the question about the fate of the asymptotic multipartite entanglement when ETH is satisfied, see Eq.(1)(main text). 
For a sufficiently chaotic Hamiltonian, a generic initial state  $\ket{\psi}$ will be such that $|c_n|^2$ has a narrow distribution around an average energy
$  E  = \bra {\psi} \hat H \ket{\psi}$, 
meaning that the energy fluctuations in the diagonal ensemble 
 $\delta^2 E = \bra {\psi} \hat H^2 \ket{\psi} - \bra {\psi} \hat H \ket{\psi}^2$
are sufficiently small, i.e. $\delta E^2 / E^2 \sim 1/N$.
Hence, the fluctuations of the operator $\hat O$ over the diagonal ensemble 
\begin{equation}
\aveDG{\Delta^2 \hat O} 
     = \sum_n\, |c_n|^2 \, (O_{nn})^2 -\left ( \sum_n\, |c_n|^2 \, O_{nn} \right )^2\ , 
\end{equation}
can be evaluated by a Taylor expansion of diagonal smooth function of the ETH around the mean energy
\begin{equation}
\label{eq:taylor}
O_{nn} = O(E) + (E_n-E) \, \left.\pd{O}{E_n}\right|_E + \frac 12 \, (E_n-E)^2\,\left.{\frac{\partial^2 O}{\partial E_n^2}}\right|_E + \dots  \ .
\end{equation}
Substituting back, keeping terms up to $\mathcal{O}(\delta E^2)$ one obtains \cite{Alessio:2016}

\begin{align}
\label{eq:flu_diag}
\aveDG{\Delta^2 \hat O} 
   & = \bra{E_n} \Delta^2 \hat O \ket{E_n} +  \left ( \frac{\partial O}{\partial E_n} \right )^2 \, \delta^2 E \ ,
\end{align}
where $\ket{E_n}$ is the eigenstate corresponding to the mean energy $E=E_n$. The fluctuations of any observable in the diagonal ensemble have essentially two independent contributions, the first coming from fluctuations within each eigenstate and the second from the  energy fluctuations.
For intensive observables, the second contribution becomes subleading ({since $ \bra{E_n} \Delta^2 \hat O \ket{E_n}\sim 1$, $\delta^2 E / E^2 \sim 1/N$ and $O'\sim 1$}). 
For extensive observables -- relevant for the QFI -- these two contributions are of the same order ({since  $ \bra{E_n} \Delta^2 \hat O \ket{E_n}\sim N$, $\delta^2 E / E^2 \sim 1/N$ and $O'\sim N$}) and fluctuations between different eigenstates might become relevant.

Finally, substituting back into Eq.~\eqref{eq:asy_FQ} one obtains Eq.~(9)(main text), which is one of our main results. The multipartite entanglement of the asymptotic state for quenched dynamics is always bounded by the contribution of the single eigenstate corresponding to the initial energy. 
%


 \subsection{Derivation of FDT for a single eigenstate}
 In this section, we review the derivation of the FDT for a single eigenstate (cf. Eqs.(2-3) of the main text), following Ref.~\ocite{Alessio:2016}.
 Consider an energy eigenstate $\ket{E_n}$, such that $E_n=\bra{E_n} \hat H \ket{E_n}$. Let us first focus on the symmetric correlation function 
\begin{widetext}
\begin{align}
\label{eq:S_singlestate}
S_{\hat O}(E_n,t) 
    & = 
    \bra{E_n} \{\hat O(t), \hat O(0)\}\ket{E_n}- 2\bra{E_n} \hat O(t) \ket{E_n} \, \bra{E_n}\hat O(0) \ket{E_n} \nonumber \\
&      = 
     \sum_{m: m\neq n} \left (e^{i\, \omega_{nm}t} + e^{- i\, \omega_{nm}t} \right) |R_{nm}|^2\,      
|f_{\hat O}(E_n+\omega_{mn}/2, \omega_{nm})|^2 e^{S(E_n+\omega_{mn}/2)} \ ,    
\end{align}
\end{widetext}
where we have used the matrix elements of ETH (cf. Eq. (1) of the main text) and $E_{nm} = E_n + \omega_{nm}/2$, ${E_{n}=E_m+\omega_{nm}}$. We now replace $|R_{nm}|^2$ with its statistical average (unit) and each sum as an integral with the suitable density of states, $\sum_m \to \int_{0}^{\infty} dE_m\, e^{S(E_m)}= \int d\omega e^{S(E+\omega)}$. Thus,
\begin{align}
\begin{split}
S_{\hat O}&(E_n,t) = \\
&\int d{\omega}\, (e^{i\omega t} + e^{-i \omega t} ) |f_{\hat O}(E_n+\omega/2, \omega)|^2e^{S(E_n+\omega/2)} \ .
\end{split}    
\end{align}
Since $f_{\hat{O}}(E, \omega)$ decays rapidly enough at large $\omega$ ~\cite{murthy2019}, we can expand the exponent in powers of $\omega$

\begin{subequations}
\begin{align}
S(E_n+\omega) - S(E_n+\omega/2) & 
    = \frac {\beta \omega}2 -\frac 38 \, \omega^2 \,  \frac {\beta^2}C + \dots \ , \\
f_{\hat O}(E_n+\omega/2, \omega) & = f_{\hat O}(E_n, \omega) + \pd{f_{\hat O}}{E_n} \, \frac {\omega} 2 + \dots
\end{align}
\end{subequations}
where in the first line we have substituted the thermodynamic definitions of temperature $\beta = S'(E)$ and of heat capacity $S''(E) = -\beta^2 /C$. In the thermodynamic limit also the (extensive) heat capacity $C$ diverges, hence one can neglect $S''$. 
Substituting back, we obtain
\begin{widetext}
\begin{align}
\begin{split}
S_{\hat O}(E_n,t)& =
\int d{\omega}\, (e^{i\omega t} + e^{-i \omega t} ) 
 |f_{\hat O}(E_n, \omega + \omega / 2)|^2   \\
     & = 2 
\int d{\omega}\,   e^{i\omega t}\, 
\left [
|f_{\hat O}(E_n, \omega)|^2 \, \cosh{\beta \omega /2}+ \frac {\omega}2\, \pd{|f_{\hat O}(E_n, \omega)|^2 }{E} \sinh{\beta \omega /2}
\right ] \ .
\end{split}
\end{align}
The ETH expression for the symmetric response function reads
\begin{align}
\begin{split}
\label{eq:S_Enmega}
& S_{\hat O}(E_n, \omega) = 4
 \pi\, \left [
|f_{\hat O}(E_n, \omega)|^2 \, \cosh{\beta \omega /2}+  \frac {\omega}2\, \pd{|f_{\hat O}(E_n, \omega)|^2 }{E} \sinh{\beta \omega /2}
\right ] \ .
\end{split}
\end{align}
\end{widetext}
Notice that, if $\hat O$ is a local operator, or a sum of local operators, the term containing the energy derivative becomes irrelevant in the thermodynamic limit, leading to Eq.~(\tc{2}) in the main text. \\

We can now repeat all the same calculations for the Kubo susceptibility (\ref{eq:chi_def}). Substituting the matrix elements suggested by ETH in the susceptibility we obtain
\begin{widetext}
\begin{align}
\begin{split}
\chi_{\hat O}(E_n,t)  
       & = - i \theta(t) \, \sum_{m: m\neq n} \left (e^{i\, \omega_{nm}t} - e^{- i\, \omega_{nm}t} \right) |R_{nm}|^2\,
       |f_{\hat O}(E_n+\omega_{mn}/2, \omega_{nm})|^2 e^{S(E_n+\omega_{mn}/2)} \ .
\end{split}    
\end{align}
By substituting the average of the fluctuating part and replacing sums with the corresponding integrals, we obtain
\begin{align}
\begin{split}
& \chi_{\hat O}(E_n,t)   =
-2\, i\,  \theta(t) \int d{\omega}\,
\left [
|f_{\hat O}(E_n, \omega)|^2 \, \sinh{\beta \omega /2}+ \frac {\omega}2\, \pd{|f_{\hat O}(E_n, \omega)|^2 }{E} \cosh{\beta \omega /2}
\right ] \ .
\end{split}    
\end{align}
If we now take the Fourier transform $-i\int_{-\infty}^{\infty} \theta(t)e^{ixt} = 1/(x+i0^+) = \mathcal P / x - i \pi \delta(x)$, 
%
%
the ETH expression for the Kubo susceptibility [$\chi''_{\hat O}(\omega)= - \text{Im}\chi_{\hat O}(\omega)$] reads
\begin{align}
\begin{split}
\label{eq:chi_Enmega}
& \chi_{\hat O}''(E_n, \omega)  = 
 2 \pi \, \left [
|f_{\hat{O}}(E_n, \omega)|^2 \, \sinh{\beta \omega /2}+ \frac {\omega}2\, \pd{|f_{\hat{O}}(E_n, \omega)|^2 }{E} \cosh{\beta \omega /2}
\right ] \ ,
\end{split}
\end{align}
\end{widetext}
which leads to Eq.~(2) of the main text.

\medskip

\subsection{FDT for the diagonal ensemble}
Le us now see how this picture is modified when we consider the expectation values over the diagonal ensemble from Eq.~\eqref{eq:DE}. We start by considering the symmetric fluctuation from Eq.~\eqref{eq:corre_fun_DE_S}. By expanding in the energy eigenbasis and considering diagonal and non-diagonal terms, one obtains
\begin{align}
\begin{split}
& S_{\hat O}^{\text{DE}}(E, t) =
 \sum_n\, |c_n|^2\, S_{\hat O}(E_n, t) 
   +2\, 
\left ( \frac{\partial O}{\partial E_n} \right )^2 \, \delta^2 E
 \ .
\end{split}    
\end{align}
The first term represents the average over the diagonal ensemble of the single eigenstate contribution $S_{\hat O}^{\text{DE}}(E_n, t) $ in Eq.~\eqref{eq:S_singlestate}, that we discussed in the previous Section. The second contribution is the the same as the one appearing for the static fluctuations in Eq.~\eqref{eq:flu_diag} and it is derived in the same way, by expanding the diagonal matrix elements.
Analogously, for the linear response function
\begin{equation}
    \chi_{\hat{O}}^{\text{DE}}(E, t) = \sum_n\, |c_n|^2\, \chi_{\hat O}(E_n, t).
\end{equation}
It is now clear that by expanding around the average value $E$ as in Eq.~\eqref{eq:taylor}, we obtain another second order contribution
\begin{align}
\label{eq:S_correction}
S_{\hat O}^{\text{DE}}(E, \omega) &= S_{\hat O}(E_n, \omega) \nonumber \\
&+ \delta E^2 \,\left [ \frac{\partial^2 S_{\hat O}(E_n, \omega)  }{\partial E_n^2}   +  \left ( \pd{O}{E_n} \right ) ^2 
    \right], \nonumber \\
\chi_{\hat O}^{\prime \prime\,\text{DE}}(E, \omega) & = \chi''_{\hat O}(E_n, \omega) + \delta E^2  \,\left [ \frac{\partial^2 \chi''_{\hat O}(E_n, \omega)  }{\partial E_n^2} \right ],
\end{align}
where the correction can be directly evaluated from Eqs.~\eqref{eq:S_Enmega}-\eqref{eq:chi_Enmega}, extracting the smooth function $f_{\hat{O}}(E, \omega)$. However, as discussed before, the contribution of these derivatives should vanish in the thermodynamic limit, yielding the the same results in Eqs.~(2-3) of the main text.

\section{Quantum Fisher information within ETH}
\label{sec:QFI_ETH}
In this section we will present an alternative derivation of Eq.~(6) in the main text, which is based only on ETH, without the assumption of a Gibbs distribution \cite{Hauke2016}. The QFI, defined in Eq.~(5) in the main text, can be written as
\begin{align}
\label{eq:QFI_sum}
    \begin{split}
        \mathcal F (\hat O) & = \sum_n\, p_n\, \mathcal F_{\text{ETH}}(\hat O) \, - 8 \, \sum_{n\neq m}\, \frac{p_n\, p_m}{p_n+p_m} |O_{nm}|^2 \ ,
    \end{split}
\end{align}
where $\mathcal F_{\text{ETH}} = 4\, \langle E_n | \Delta \hat{O}^2 |E_ n \rangle$ is the QFI for a single energy eigenstate [cf. Eq.~(7) of the main text]. \\
Consider a generic density matrix in the energy eigenbasis 
with $p_n=p(E_n)$ smooth functions of the energy eigenvalue $E_n$. For example, $\hat \rho$ can represent either the canonical Gibbs distribution ($p_n=e^{-\beta\, E_n} / Z$) or the micro-canonical one  (MC) ($p_n$ constant on a small energy shell of width $\delta E$). 

Let us start with the first term in Eq.~\eqref{eq:QFI_sum}. By replacing the sum with an integral with the suitable density of states, $\sum_n \to \int_{0}^{\infty} dE'\, e^{S(E')}$, we obtain
\begin{equation}
\label{eq:eth_singleF}
    \int_0^{\infty} {d}E'\, p(E') \, e^{S(E')} \, \mathcal F_{\textrm{pure}}(E') \ . 
\end{equation}

The second term in Eq.~\eqref{eq:QFI_sum} vanishes in the case of a pure eigenstate $\ket{E_n}$ , i.e. $p_m=\delta_{nm}$, and the QFI is given by Eq.(7) of the main text. On the other hand, in the case of  a thermodynamic ensemble, using the ETH ansatz [cf. Eq.(1) of the main text],  we can re-write it as
\begin{widetext}
\begin{align}
\label{eq:palla}
\begin{split}
\sum_{n\neq m} \frac{p_mp_n}{p_m+p_n} |O_{nm}|^2  
    & =
    \sum_{n\neq m} \frac{p_mp_n}{p_m+p_n} |R_{nm}|^2 \, |f_{\hat O}(E_{nm}, \omega_{mn})|^2\, e^{-S(E_{nm})}  \\
        & = \int_0^{\infty} dE' \int_{-\infty}^{\infty} \, {d}\omega \, \frac{p(E'+\omega / 2) p(E'-\omega / 2)}{p(E'+\omega / 2)+p(E'-\omega / 2)}  \,|f_{\hat O}(E', \omega)|^2\, 
        e^{-S(E') + S(E'-\omega / 2) + S(E'+\omega / 2)} \ ,
\end{split}
\end{align}
where in the second line we have replaced $|R_{nm}|^2$ with its statistical average $1$, $E_{n,m}= E\pm \omega / 2$ and  each sum as an integral with the suitable density of states, $\sum_{n} \sum_n \to \int_{0}^{\infty} dE' \int_{-\infty}^{\infty} d\omega \, e^{S(E'+\omega /2)+ S(E'-\omega / 2)}$. Since $f_{\hat{O}}(E', \omega)$ decays rapidly enough at large $\omega$ ~\cite{murthy2019}, we can expand the exponent in powers of $\omega$
\begin{align}
-S(E') + S(E'+\omega/2) + S(E'-\omega / 2)  & = S(E') - \left( \frac{\beta \omega}2\right )^2 \, \frac 1C + \dots
\end{align}
\end{widetext}
where $C$ is the (extensive) heat capacity of the system defined above. Hence, as already done in the previous section, we will ignore its contribution in the thermodynamic limit. We now consider the ratio between the different probabilities that, as we will show at the end of the section, for the MC and Gibbs ensembles factorizes as
\begin{equation}
\label{eq:come}
 \frac{p(E'+\omega / 2) p(E'-\omega / 2)}{p(E'+\omega / 2)+p(E'-\omega / 2)}  = \frac 12\, \frac{p(E')}{\cosh(\beta \omega / 2)}  \ ,
\end{equation}
where $\beta = \partial S(E)/\partial E|_{E'}$. Substituting back into Eq.~\eqref{eq:palla} we obtain
\begin{align}
    \begin{split}
   & \frac 12 \int \text{d}E'\,p(E') e^{S(E')}\, \int \text{d}\omega\, |f_{\hat O}(E', \omega)|^2\, \frac{  1}{\cosh(\beta \omega / 2)} \\
   & = \frac 1{8 \pi} \, \int \text{d}E'\,p(E') e^{S(E')}\, \int \text{d}\omega\, \frac{  S_{\hat O}(E', \omega)}{\cosh^2(\beta \omega / 2)}
\end{split}
\end{align}
where in the second line we have used the ETH expression of the symmetrized noise [cf. Eq.(3) in the main text]. This equation together with Eq.~\eqref{eq:eth_singleF} and  Eq.~\eqref{eq:QFI_sum} yields
\begin{align}
    \begin{split}
        \mathcal{F}(\hat O)
        = \frac 1{\pi}  \int \text{d}E'\,p(E') e^{S(E')}\, \int \text{d}\omega\,  S_{\hat O}(E', \omega)\tanh^2(\beta \omega / 2) \ ,
    \end{split}
\end{align}
where we have employed $1 - \, \frac 1{\cosh^2(\beta \omega / 2)} = \tanh^2(\beta \omega /2)$. 
The energy integral can be evaluated explicitly depending on the ensemble. In the case of the Gibbs distribution,
the integral can now be solved in the thermodynamic limit via the saddle point approximation, yielding the same result of $\mathcal F_{\text{Gibbs}}$ as in Eq.~(6) of the main text.  For the MC, one has the uniform average of the same expression over the energy shell. \\

Let us now prove Eq.~\eqref{eq:come}. 
In the case the canonical distribution we have 
\begin{align}
 \frac{p(E'+\omega / 2) p(E'-\omega / 2)}{p(E'+\omega / 2)+p(E'-\omega / 2)} \nonumber 
& =\frac{e^{-2 \beta E'}}{Z\, e^{-\beta E'} \, (e^{-\beta \omega / 2} + e^{\beta \omega / 2}) } \\
    & = \frac 12\, \frac {p(E')}{\cosh{\beta \omega /2}}\ .\nonumber
\end{align}
On the other hand, let us consider the MC distribution 
\[
p(E') \sim  
\begin{dcases}
 e^{-S(E')}\, \delta E^{-1} \quad \forall \; E \in [{E}, {E} + \delta E] \\
 0 \quad \quad \quad \quad \quad \quad \text{otherwise}
\end{dcases} \ .
\]
%
by expanding in powers of $\omega$ one finds
\begin{align*}
 \frac{p(E'+\omega / 2) p(E'-\omega / 2)}{p(E'+\omega / 2)+p(E'-\omega / 2)} & =
\frac{p^2+ (\omega/2)^2\, \left [ p\, p^{\prime \prime} - (p^{\prime})^2 \right ]} {2p + (\omega / 2)^2\, p^{\prime \prime}} \\
& = \frac 12 \, \frac {p(E')}{1+ (\beta \omega )^2/2} \ ,
\end{align*}
where we have used $p=p(E')$ and $p^{\prime} = \partial p(E) / \partial E|_{E'} = -S^{\prime} p(E') = -\beta p(E') $ and $p^{\prime \prime} = \partial^2 p(E') / \partial E'^2 = \beta^2 p(E') $, since as already discussed, $S^{\prime \prime}$ vanishes in the thermodynamic limit. It is now easy to see that this equation corresponds to the result in Eq.~\eqref{eq:come}, at the leading orders in $\omega$. Notice that this expansion can in principle be generalized to other smooth distributions $p(E')$.\\

\section{Experimental Consequences}
\label{sec:experiment}

In the following we discuss a possible scheme to appreciate the consequences of our result experimentally. Since ETH applies to chaotic quantum systems undergoing unitary evolution, then the type of experimental setup would have to be extremely well isolated from the laboratory environment. Specifically one needs the thermalisation time $\tau_{th} \ll \tau_{\phi}$, where $\tau_{\phi}$ is the coherence time, this condition is routinely achieved in ultra-cold atomic physics. In these experiments, one would prepare a pure state, which could be the ground state of some simulated Hamiltonian. Such a pure state preparation is routinely performed in these setups. A quench is then performed and the state of the system -- represented by a density matrix $\rho(t)$ -- is a superposition of energy eigenstates with an energy which is conserved in the subsequent evolution. In order to extract the correct QFI one could try to certify the purity of the state during time evolution following quench. Performing this procedure beyond full state tomography is challenging. Under these assumptions, the correct ETH value of the QFI could be obtained by just measuring the fluctuations of observables such as the density.

However, a possible experimental signature could be obtained by an alternative procedure.  Consider the time evolution of a generic ergodic many-body system with $\tau_{th} \ll \tau_{\phi}$. If we let the system evolve up to time $t$ and use $\rho(t)$ for \emph{phase estimation purposes}\cite{Pezz2018}, i.e. to estimate an unknown phase $\phi$. The accuracy of this protocol $\Delta \phi(t)$ is related to the time-dependent QFI via the quantum Cramer-Rao bound, i.e. $\Delta \phi(t)\leq 1/\sqrt{M \mathcal F(O, t)}$, where $M$ is the number of measurements~\cite{paris2009quantum}.
Our result has direct consequences for $\Delta \phi(t)$ for $t>\tau_{th}$. 
In fact, as long as the state is pure, $\Delta \phi(t)$ will reach a constant value, which is bounded by the ETH result. \\
In turn, as time $t$ gets beyond $\tau_{\phi}$, the maximum accuracy will increasingly degrade, since the state will get progressively mixed and the Gibbs result will hold.
It is natural to suspect that in certain conditions, this degradation could be studied in detail. This follows from the fact that, e.g., close to thermal transitions the ETH QFI is diverging while the other stays finite. This might be hard to check experimentally, but a priori it is possible and it would yield a very interesting proof of principle.

\bibliography{ethfisher}